# B.1.258Δ, a SARS-CoV-2 variant with ΔH69/ΔV70 in the Spike protein circulating in the Czech Republic and Slovakia

Broňa Brejová[1,*], Viktória Hodorová[2], Kristína Boršová[3,4], Viktória Čabanová[3], Lenka Reizigová[5,6], Evan D. Paul[7], Pavol Čekan[7], Boris Klempa[3], Jozef Nosek[2], Tomáš Vinař[1,*]

**Affiliations:**

[1] Faculty of Mathematics, Physics and Informatics, Comenius University in Bratislava, Mlynská dolina, 842 48 Bratislava, Slovak Republic
[2] Department of Biochemistry, Faculty of Natural Sciences, Comenius University in Bratislava, Ilkovičova 6, 842 15 Bratislava, Slovak Republic
[3] Institute of Virology, Biomedical Research Center of the Slovak Academy of Sciences, Dúbravská cesta 9, 845 05 Bratislava, Slovak Republic
[4] Department of Microbiology and Virology, Faculty of Natural Sciences, Comenius University in Bratislava, Ilkovičova 6, 842 15 Bratislava, Slovak Republic
[5] Regional Authority of Public Health, Trenčín, Slovakia
[6] Department of Laboratory Medicine, Faculty of Healthcare and Social Work, Trnava University, Trnava, Slovakia
[7] MultiplexDX, Bratislava, Slovakia

***Address for correspondence:**
Faculty of Mathematics, Physics and Informatics, Comenius University in Bratislava, Mlynská dolina, 842 48 Bratislava, Slovak Republic; brejova@dcs.fmph.uniba.sk; tomas.vinar@fmph.uniba.sk

## Summary

SARS-CoV-2 mutants carrying the ΔH69/ΔV70 deletion in the amino terminal domain of the Spike protein emerged independently in at least six lineages of the virus (namely, B.1.1.7, B.1.1.298, B.1.160, B.1.177, B.1.258, B.1.375). Routine RT-qPCR tests including TaqPath or similar assays based on a drop-out of the Spike gene target are incapable of distinguishing among these lineages and often lead to the false conclusion that clinical samples contain the B.1.1.7 variant, which recently emerged in the United Kingdom and is quickly spreading through the human population. We analyzed SARS-CoV-2 samples collected from various regions of Slovakia between November and December 2020 that were presumed to contain the B.1.1.7 variant due to traveling history of the virus carriers or their contacts. Sequencing of these isolates revealed that although in some cases the samples were indeed confirmed as B.1.1.7, a substantial fraction of isolates contained another ΔH69/ΔV70 carrying mutant belonging to the lineage B.1.258, which has been circulating in Central Europe since August 2020, long before the import of B.1.1.7. Phylogenetic analysis shows that the early sublineage of B.1.258 acquired the N439K substitution in the receptor binding domain (RBD) of the Spike protein and, later on, also the deletion ΔH69/ΔV70 in the Spike N-terminal domain (NTD). This variant is particularly common in several European countries including Czech Republic and Slovakia, and we propose to name it B.1.258Δ.

## Introduction

Monitoring clinical samples by rapid and near real-time genome sequencing of the SARS-CoV-2 virus provides an important means for policy decisions in combating the COVID-19 pandemic (Oude Munnink et al. 2020). In particular, the emergence of new virus variants exhibiting enhanced infectivity such as B.1.1.7 (20I/501Y.V1), B.1.351 (20H/501Y.V2), and P.1 (20J/501Y.V3) recently detected in the United Kingdom, South Africa, and Brazil, respectively, increased the attention of national public health authorities (Rambaut et al. 2020a; Tegally et al. 2020; Kosakovsky Pond et al. 2020; NIoID 2021; Faria et al. 2021; Naveca et al. 2021a,b).

Here, we describe a variant named B.1.258Δ carrying the ΔH69/ΔV70 mutation in the Spike protein that has emerged within the B.1.258 clade and has been prevalent in the Czech Republic (approx. 59% out of 251 sequenced samples between September and December 2020), Slovakia (25% out of 72 sequenced samples in the same months), and several other countries **(Table 1)**. The ΔH69/ΔV70 deletion in the Spike N-terminal domain (NTD) is associated with increased infectivity and evasion of the immune response (Kemp et al. 2020a) and evidence suggests that this mutation has arisen in B.1.258 independently of the B.1.1.7 variant. The deletion is likely to cause a drop-out of the Spike gene target in TaqPath RT-qPCR (Volz et al. 2020; Washington et al. 2020; Bal et al. 2021; Borges et al. 2021) and other assays targeting this deletion, and thus its carriers can be easily misidentified as B.1.1.7 (as it recently happened in Slovakia, see PHAoSR, 2021).

The B.1.258Δ variant also contains the N439K mutation in the receptor binding domain (RBD) of the Spike protein, likely preceding the emergence of the ΔH69/ΔV70 mutation in this lineage. The N439K substitution enhances the binding affinity to the angiotensin-converting enzyme 2 (ACE2) receptor and has been shown to facilitate immune escape from a panel of neutralizing monoclonal antibodies, as well as from polyclonal sera from persons recovered from the infection. This mutation is also associated with slightly higher viral loads (Thomson 2021). The deletion ΔH69/ΔV70 emerged recurrently in diverse lineages and it frequently co-occurs with mutations in the Spike RBD such as N439K, Y453F, and N501Y (Kemp et al. 2020b; Lassaunière et al. 2020; McCarthy et al. 2020; Rambaut et al. 2020; Bazykin et al. 2021; Larsen et al. 2021). It has been demonstrated that ΔH69/ΔV70 enhances the virus infectivity by two-folds in a pseudotyping assay and also compensates for mutations in the Spike RBD that lower the infectivity (*e.g.* N501Y). In addition, a non-RBD binding monoclonal antibody is less potent against the ΔH69/ΔV70 mutant (Kemp et al. 2020a). Other non-synonymous mutations characteristic for most of the B.1.258Δ samples result in amino acid substitutions in proteins involved in virus replication, *i.e.* M101I, localized in the dimerization motif 'GXXXG' (100-104) of RNA-binding protein NSP9, V720I in the palm subdomain of RNA-dependent RNA polymerase NSP12, and A598S near the C-terminus of helicase NSP13 (Zhang et al. 2020; Wang et al. 2020).

The association of mutations in the B.1.258Δ variant with increased virulence and evasion of immune responses taken together with the high prevalence in several countries (notably the Czech Republic, currently reporting one of the highest incidences of new cases per 100,000 population in the world) warrants further investigation of this variant.

## Results and Discussion

The ΔH69/ΔV70 mutation has arisen independently at least six times in different SARS-CoV-2 lineages (**Figure 1**). Besides the B.1.258Δ and B.1.1.7 variants, the mutation has also been observed in B.1.1.298 (Denmark outbreak associated with mink farms; Lassaunière et al. 2020; Larsen et al. 2021) and B.1.375 (a clade originating in the United States; Larsen and Worobey, 2020; Moreno et al. 2021). Interestingly, the ΔH69/ΔV70 mutation has recently emerged within previously characterized clades EU1 (B.1.177) and EU2 (B.1.160). This recurrent emergence even within well-established clades supports the hypothesis that ΔH69/ΔV70 mutation can compensate for other mutations that would by themselves lower the infectivity and in concert with those other mutations can lead to new variants with increased fitness and potential to evade the immune system response (Kemp et al. 2020b).

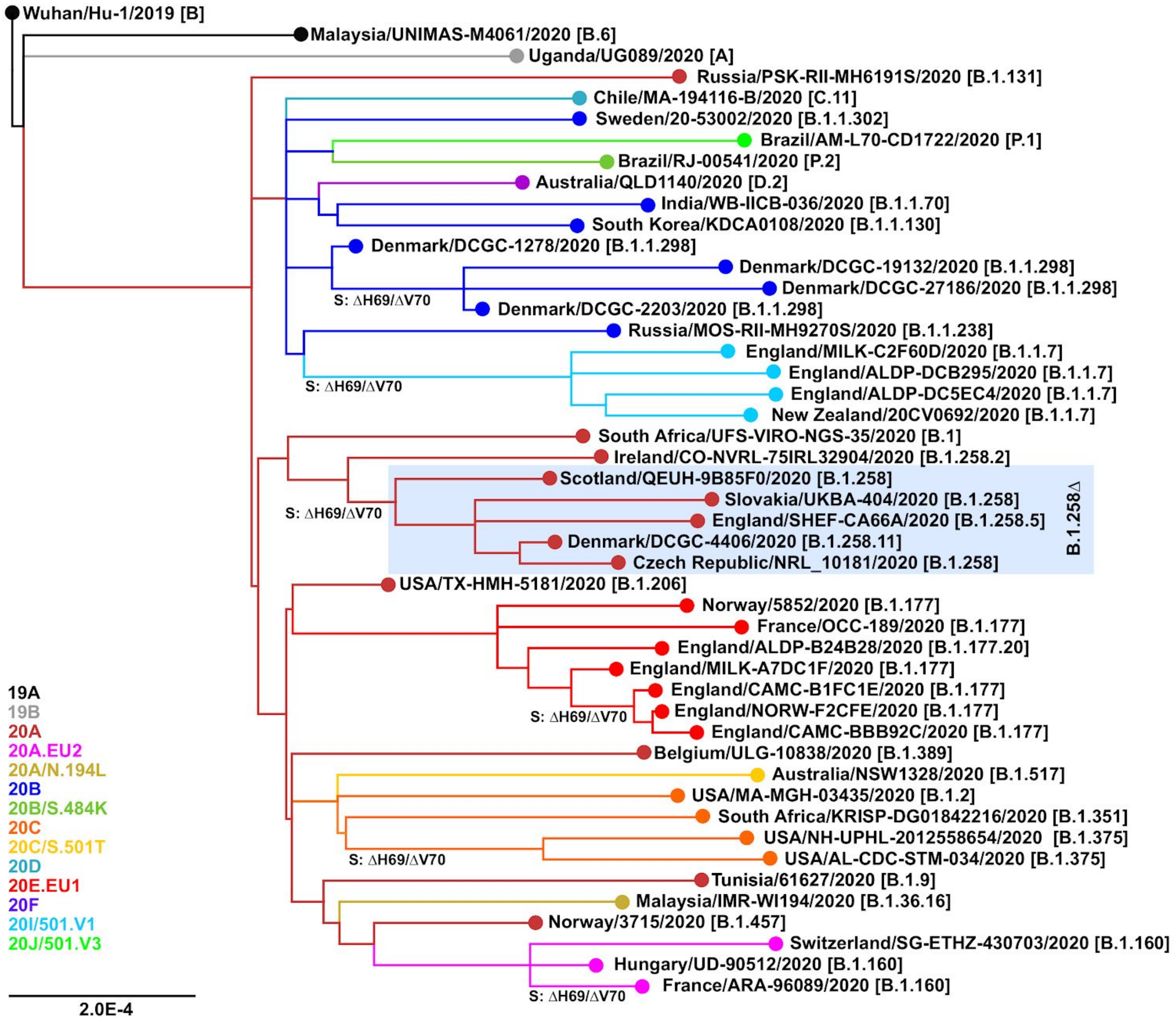

**Figure 1. Recurrent emergence of the ΔH69/ΔV70 mutation.** The points of emergence of the ΔH69/ΔV70 deletion are marked. Nextclade (https://clades.nextstrain.org; Hadfield et al. 2018) assignments are shown in color. Pangolin lineages (https://pangolin.cog-uk.io; Rambaut et al., 2020b) are indicated in brackets.

The earliest samples of the B.1.258Δ variant have been observed in Switzerland and in the United Kingdom at the beginning of August 2020 (**Figure 2**). The outgroup sample that already contains the Spike N439K mutation, but not the ΔH69/ΔV70 deletion, has been sampled in Romania as early as May 13, 2020; before that, the outgroup that does not contain N439K has been observed in England on March 22, 2020 (EPI_ISL_423656). Subsequently, the B.1.258Δ variant has spread and gained significant prevalence in multiple countries, including the Czech Republic, Sweden, Slovakia, Poland, Denmark, and Austria (**Table 1**). Note that clade B.1.258 also includes sequences without the ΔH69/ΔV70 deletion. In particular, most samples in sublineages B.1.258.1-B.1.258.3 and B.1.258.14-B.1.258.16 do not have the deletion, while those in B.1.258.4-B.1.258.12 and B.1.258.17-B.1.258.21 mostly have it.

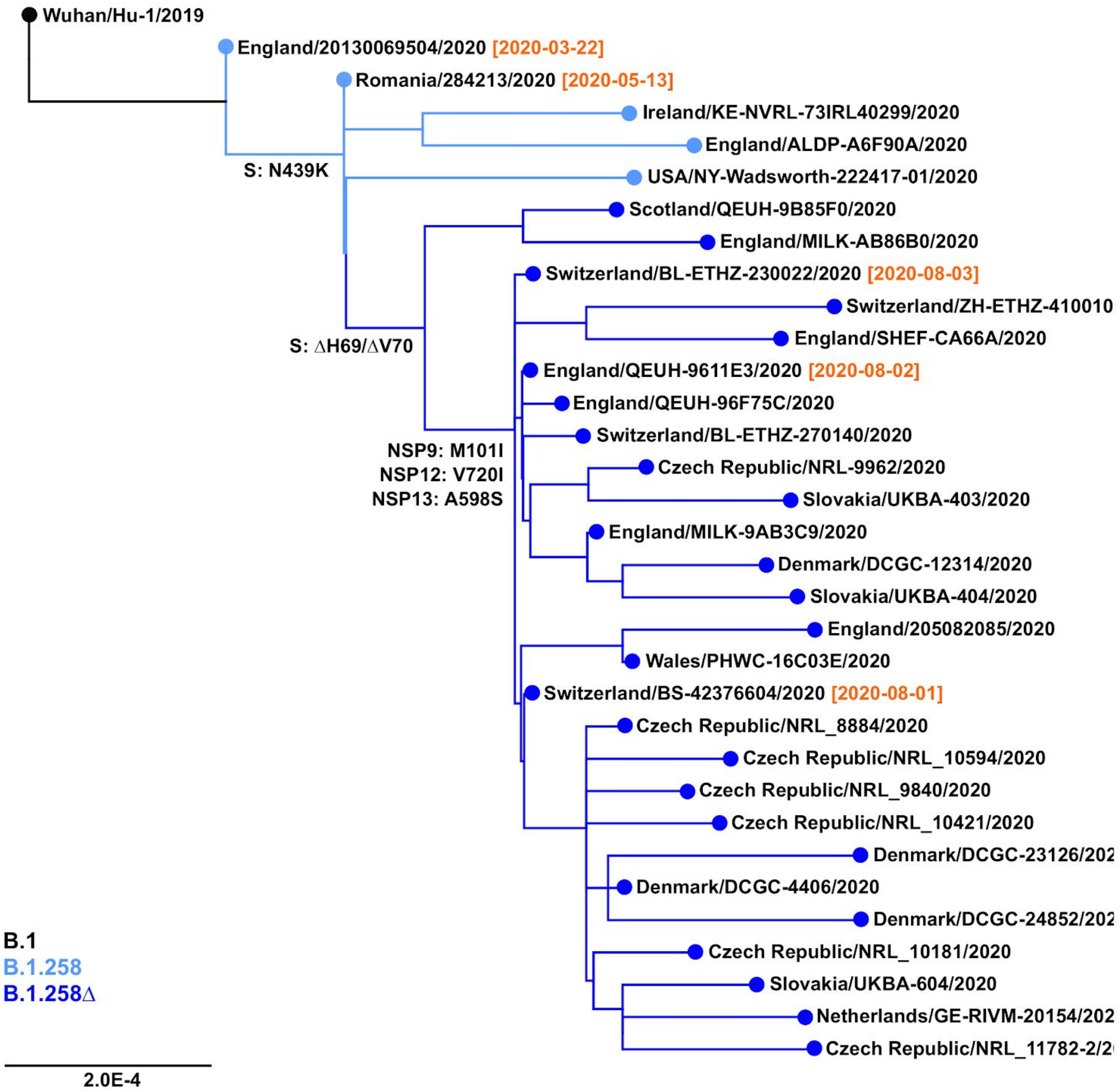

**Figure 2. Origins of B.1.258Δ variant.** Positions of mutations S:N439K, S:ΔH69/ΔV70 and additional substitutions defining the variant (*i.e.* NSP9:M101I, NSP12:V720I, NSP13:A598S) are marked. Collection dates are shown in samples nearest to the important branching points.

**Table 1. Prevalence of B.1.258Δ and B.1.1.7 variants in selected countries between September and December 2020.** The numbers represent the percentage of all samples uploaded to GISAID within the period. Only percentages based on at least 20 samples are shown, numbers based on fewer than 100 samples are shown in parentheses.

| | B.1.258Δ | | | | B.1.1.7 | | | |
|---|---|---|---|---|---|---|---|---|
| | 2020-09 | 2020-10 | 2020-11 | 2020-12 | 2020-09 | 2020-10 | 2020-11 | 2020-12 |
| **Austria** | | (0) | 22 | | | (0) | 0 | |
| **Czech Republic** | (40) | 66 | (54) | | (0) | 0 | (0) | |
| **Denmark** | 11 | 16 | 10 | 9 | 0 | 0 | <1 | 1 |
| **Germany** | 6 | 7 | 9 | 11 | 0 | 0 | <1 | 3 |
| **Iceland** | 2 | 5 | 8 | 4 | 0 | 0 | 0 | 5 |
| **Italy** | 0 | 0 | <1 | 2 | 0 | 0 | 0 | 14 |
| **Netherlands** | <1 | 4 | 3 | 8 | 0 | 0 | <1 | 7 |
| **Poland** | | | 29 | (23) | | | 0 | (1) |
| **Slovakia** | | | (19) | (38) | | | (0) | (41) |
| **Sweden** | (2) | (1) | (2) | 43 | (0) | (0) | (0) | 5 |
| **Switzerland** | 2 | 2 | 2 | 4 | 0 | 0 | <1 | 2 |
| **USA** | <1 | 0 | 0 | <1 | 0 | 0 | 0 | 1 |
| **United Kingdom** | 3 | 3 | 2 | 1 | <1 | <1 | 6 | 43 |

In addition to sequence analyses, we used a recently developed RT-qPCR assay which permits differentiation of the B.1.1.7 variant from other variants harboring only the ΔH69/ΔV70 deletion (Kovacova et al. 2021) in order to analyze a panel of 122 SARS-CoV-2 positive clinical samples collected within a mass testing campaign in the city of Trenčín (Western Slovakia) in December 2020. The assay indicated that 50/122 (41.0%) of the samples were carrying only the ΔH69/ΔV70 deletion while only 5/122 (4.1%) were identified as the B.1.1.7 lineage. Interestingly, analysis of the Ct values from the routine RT-qPCR assay showed significantly lower Ct values in the swab samples for both B.1.1.7 and B.1.258Δ samples, reflecting significantly higher viral loads in patients carrying these variants (**Figure 3**).

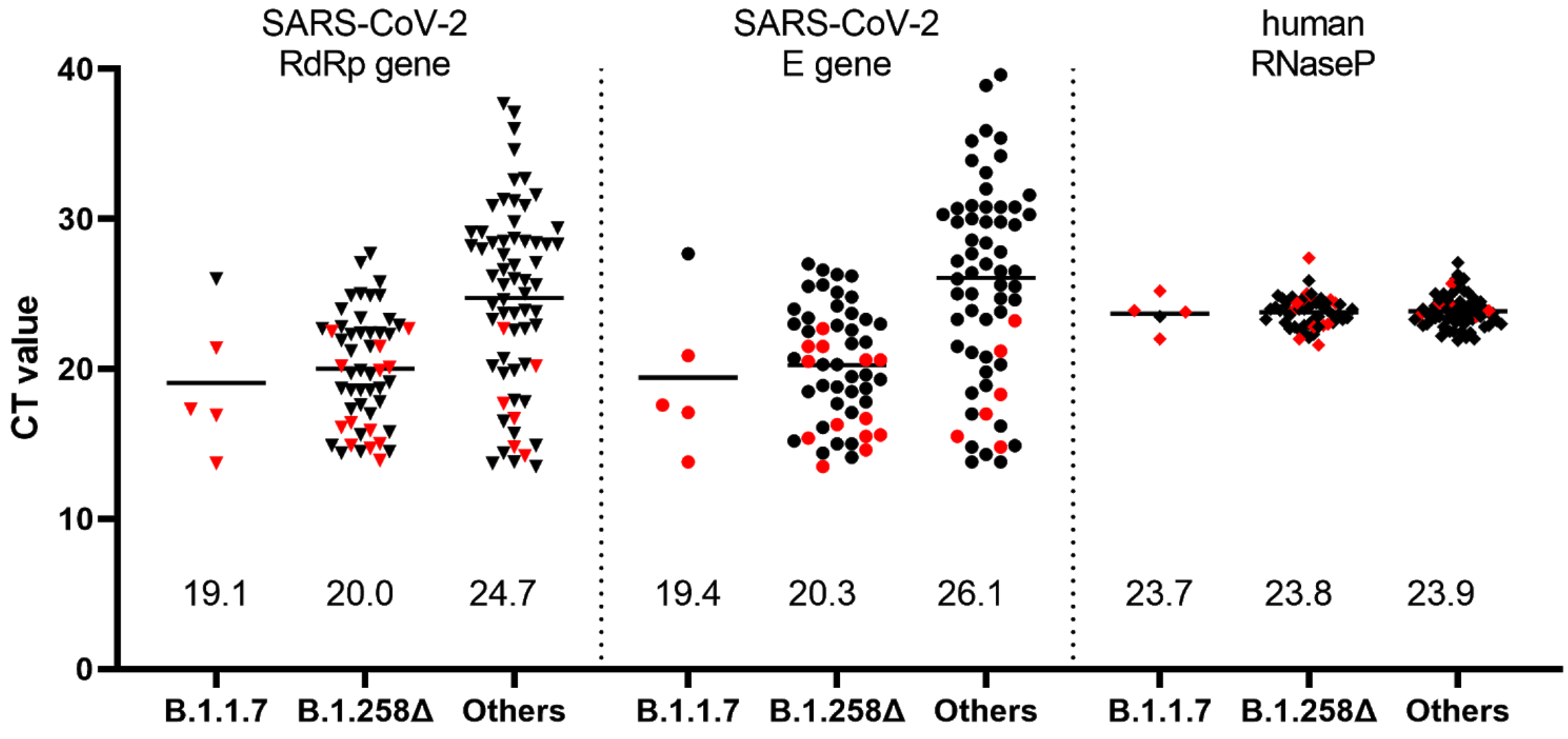

**Figure 3. Ct values in the swab specimens collected in December 2020 in the city of Trenčín (Western Slovakia) grouped according to the identified SARS-CoV-2 lineages.** The samples were initially sorted into the three groups by the RT-qPCR assay differentiating B.1.1.7 from other variants carrying ΔH69/ΔV70 deletion (Kovacova et al. 2021). The presented values are the outcome of routine SARS-CoV-2 RT-qPCR assay targeting RdRp, E, and human RNase P genes. Human RNase P assay was used as a control and indicates that the observed Ct differences are unaffected by the quality of the sample collection and RNA extraction. Proper classification of data points marked in red was further confirmed by sequencing. Numerical values shown above the x-axis indicate the mean values.

## Methods

The GISAID database was downloaded on February 3, 2021. The mutations in individual sequences were identified by comparison to the Wuhan/Hu-1/2019 sample (NCBI ID: NC_045512.2), and variant labels were assigned based on characteristic mutations identified by manual examination of the NextClade tree (for B.1.258Δ, 4 out of 5 mutations G12988T G15598A G18028T T24910C T26972C were required; for B.1.1.7 variant, 11 out of 13 mutations C3267T C5388A T6954C A23063T C23271A C23604A C23709T T24506G G24914C C27972T G28048T A28111G C28977T). The samples shown in the phylogenetic trees were selected manually to cover significant lineages of interest (B.1.258Δ variant in different countries and its outgroups, lineages containing ΔH69/ΔV70 mutation and their outgroups). Phylogenetic trees were built using Augur pipeline v. 6 (Hadfield et al. 2018). The prevalence was assessed based on all samples in GISAID for a particular country carrying particular characteristic mutations with collection dates in a particular month.

RT-qPCR assays were performed on RNA extracted by the Biomek i5 Automated Workstation using the RNAdvanced Viral kit (Beckman Coulter, Indianapolis, Indiana, USA) from swab samples collected within routine SARS-CoV-2 diagnostics (the samples were provided to researchers by health authorities without any personal identification of patients). Besides rTEST COVID-19 RT-qPCR Allplex kit (MultiplexDX, Bratislava, Slovakia) targeting the RNA-dependent RNA polymerase (RdRp) and Envelope (E) genes, the newly developed rTEST COVID-19 qPCR B.1.1.7 kit (MultiplexDX, Bratislava, Slovakia) was used to differentiate B.1.1.7 and B.1.258Δ variants (Kovacova et al. 2021). The real time PCR was performed on a QuantStudio™ 5 Real-Time PCR System (Applied Biosystem, Foster City, California, USA).

The genome sequences of SARS-CoV-2 isolates were determined on a MinION sequencer (Oxford Nanopore Technologies) using a protocol based on PCR-tiling of 2-kb long amplicons (Resende et al. 2020).

## Conclusions

We have described a B.1.258Δ variant of the SARS-CoV-2 virus that contains the S:N439K mutation shown to enhance the binding affinity of the Spike protein to human ACE2 receptor and facilitating escape from immune response, the S:ΔH69/ΔV70 mutation, which is known to increase viral infectivity, as well as several other non-synonymous mutations in NSP9, NSP12 and NSP13 likely affecting viral replication. RT-qPCR analysis on random samples collected during a mass testing campaign indicate that B.1.258Δ samples carry higher viral loads compared to other strains, similarly as in the case of the B.1.1.7 variant.

Interestingly, the B.1.258.17 sublineage of B.1.258Δ has accumulated a higher number of mutations compared to other B.1.258Δ samples, including additional substitutions in the Spike protein (L189F, V772I) and helicase NSP13 (P53L). It also has a substitution Q185H in the second B cell epitope of ORF3a protein involved in apoptosis induction (Ren et al. 2020).

Our characterization of a novel B.1.258Δ variant that has emerged in several European countries and appears to result in higher viral loads highlights the importance of vigilant genomic surveillance methods in properly identifying and tracking SARS-CoV-2 variants that display the potential to derail worldwide efforts to mitigate the pandemic.

**Acknowledgements**. We gratefully acknowledge the authors from the originating laboratories responsible for obtaining the specimens, as well as the submitting laboratories where the genome data were generated and shared via GISAID (https://www.gisaid.org/), on which this research is based (see Supplement).

**Funding.** Our research was supported by grants from the Slovak Research and Development Agency (APVV-18-0239 to JN, PP-COVID-20-0017 to BK, and PP-COVID-20-0116 to PC and BK), the Scientific Grant Agency (VEGA 1/0463/20 to BB, VEGA 1/0458/18 to TV), and the European Union's Horizon 2020 research and innovation program (EVA-GLOBAL project #871029 to BK and PANGAIA project #872539). The research was also supported in part by OPII project ITMS2014: 313011ATL7.

| GISAID Acknowledgements | | | |
|---|---|---|---|
| Originating lab | Submitting lab | Authors | Samples |
| St Vincent's Pathology (SydPath) | NSW Health Pathology - Institute of Clinical Pathology and Medical Research; Westmead Hospital; University of Sydney | CIDM-PH et al. et al | Australia/NSW1328/2020 (EPI_ISL_767894) |
| unknown | Public Health Virology Laboratory, Forensic and Scientific Services (PHV-FSS) | Son Nguyen et al. et al | Australia/QLD1140/2020 (EPI_ISL_593641) |
| Department of Clinical Microbiology | GIGA Medical Genomics | Keith Durkin et al | Belgium/ULG-10838/2020 (EPI_ISL_666873) |
| DB Diagnosticos do Brasil | Laboratório de Parasitologia Médica - Instituto de Medicina Tropical - Universidade de São Paulo | Nuno Faria et al | Brazil/AM-L70-CD1722/2020 (EPI_ISL_804823) |
| Laboratorio de Virologia Molecular / UFRJ | Bioinformatics Laboratory / LNCC | Carolina M Voloch et al | Brazil/RJ-00541/2020 (EPI_ISL_717925) |
| Genetica Molecular and Subdepartamento de Virologia ISP Chile | Instituto de Salud Publica de Chile | Javier Tognarelli et al | Chile/MA-194116-B/2020 (EPI_ISL_746529) |
| The National Institute of Public Health | State Veterinary Institute Prague | Nagy et al | CzechRepublic/NRL_10181/2020 (EPI_ISL_626625), CzechRepublic/NRL_10421/2020 (EPI_ISL_737012), CzechRepublic/NRL_10594/2020 (EPI_ISL_737032), CzechRepublic/NRL_11782-2/2020 (EPI_ISL_792704), CzechRepublic/NRL_8884/2020 (EPI_ISL_626577), CzechRepublic/NRL_9840/2020 (EPI_ISL_626609), CzechRepublic/NRL-9962/2020 (EPI_ISL_660554) |

| GISAID Acknowledgements | | | |
|---|---|---|---|
| Originating lab | Submitting lab | Authors | Samples |
| Department of Virus and Microbiological Special Diagnostics, Statens Serum Institut, Copenhagen, Denmark | Albertsen Lab, Department of Chemistry and Bioscience, Aalborg University, Denmark | Danish Covid-19 Genome Consortium et al | Denmark/DCGC-12314/2020 (EPI_ISL_682495), Denmark/DCGC-19132/2020 (EPI_ISL_748268), Denmark/DCGC-23126/2020 (EPI_ISL_793371), Denmark/DCGC-24852/2020 (EPI_ISL_795339), Denmark/DCGC-27186/2020 (EPI_ISL_817500) |
| Department of Virus and Microbiological Special Diagnostics, Statens Serum Institut, Denmark | Albertsen lab, Department of Chemistry and Bioscience, Aalborg University, Denmark | Danish Covid-19 Genome Consortia et al | Denmark/DCGC-1278/2020 (EPI_ISL_618071), Denmark/DCGC-2203/2020 (EPI_ISL_617504), Denmark/DCGC-4406/2020 (EPI_ISL_621827) |
| Respiratory Virus Unit, Microbiology Services Colindale, Public Health England | Respiratory Virus Unit, Microbiology Services Colindale, Public Health England | Monica Galiano et al | England/20130069504/2020 (EPI_ISL_423656) |
| Respiratory Virus Unit, National Infection Service, Public Health England | COVID-19 Genomics UK (COG-UK) Consortium | PHE Covid Sequencing Team et al | England/205082085/2020 (EPI_ISL_733592) |
| Lighthouse Lab in Alderley Park | Wellcome Sanger Institute for the COVID-19 Genomics UK (COG-UK) consortium | Jacquelyn Wynn et al | England/ALDP-A6F90A/2020 (EPI_ISL_607092) |

| GISAID Acknowledgements | | | |
|---|---|---|---|
| Originating lab | Submitting lab | Authors | Samples |
| Lighthouse Lab in Alderley Park | Wellcome Sanger Institute for the COVID-19 Genomics UK (COG-UK) Consortium | Jacquelyn Wynn et al | England/ALDP-B24B28/2020 (EPI_ISL_647005), England/ALDP-DC5EC4/2020 (EPI_ISL_822058), England/ALDP-DCB295/2020 (EPI_ISL_822172) |
| Lighthouse Lab in Cambridge | Wellcome Sanger Institute for the COVID-19 Genomics UK (COG-UK) Consortium | Rob Howes et al | England/CAMC-B1FC1E/2020 (EPI_ISL_647519) |
| Department of Pathology, University of Cambridge | COVID-19 Genomics UK (COG-UK) Consortium | Aminu S. Jahun et al | England/CAMC-BBB92C/2020 (EPI_ISL_704384) |
| Lighthouse Lab in Milton Keynes | Wellcome Sanger Institute for the COVID-19 Genomics UK (COG-UK) consortium | The Lighthouse Lab in Milton Keynes et al | England/MILK-9AB3C9/2020 (EPI_ISL_551056), England/MILK-A7DC1F/2020 (EPI_ISL_606346), England/MILK-AB86B0/2020 (EPI_ISL_623362) |
| Lighthouse Lab in Milton Keynes | Wellcome Sanger Institute for the COVID-19 Genomics UK (COG-UK) Consortium | The Lighthouse Lab in Milton Keynes et al | England/MILK-C2F60D/2020 (EPI_ISL_782957) |
| Quadram Institute Bioscience | COVID-19 Genomics UK (COG-UK) Consortium | Dave J. Baker et al | England/NORW-F2CFE/2020 (EPI_ISL_650120) |
| Lighthouse Lab in Glasgow | Wellcome Sanger Institute for the COVID-19 Genomics UK (COG-UK) consortium | Harper VanSteenhouse et al | England/QEUH-9611E3/2020 (EPI_ISL_531939), England/QEUH-96F75C/2020 (EPI_ISL_540238), Scotland/QEUH-9B85F0/2020 (EPI_ISL_568224) |

| GISAID Acknowledgements | | | |
|---|---|---|---|
| Originating lab | Submitting lab | Authors | Samples |
| Virology Department, Sheffield Teaching Hospitals NHS Foundation Trust/Department of Infection, Immunity and Cardiovascular Disease, The Medical School, University of Sheffield | COVID-19 Genomics UK (COG-UK) Consortium | Thushan de Silva et al | England/SHEF-CA66A/2020 (EPI_ISL_680168) |
| CNR Virus des Infections Respiratoires - France SUD | CNR Virus des Infections Respiratoires - France SUD | Antonin Bal et al | France/ARA-96089/2020 (EPI_ISL_732701) |
| CHU Purpan - Laboratoire de Virologie - Institut Fédératif de Biologie | CHU Purpan - Laboratoire de Virologie - Institut Fédératif de Biologie | Latour J. et al | France/OCC-189/2020 (EPI_ISL_804373) |
| University of Debrecen, Department of Medical Microbiology | National Laboratory of Virology, Szentágothai Research Centre | Endre Gábor Tóth et al | Hungary/UD-90512/2020 (EPI_ISL_671473) |
| The National University Hospital of Iceland | deCODE genetics | Daniel F Gudbjartsson et al | Iceland/352/2020 (EPI_ISL_424376) |
| CSIR-Indian Institute of Chemical Biology, MEDICA Supercpecialty Hospital Kolkata | CSIR-Indian Institute of Chemical Biology, MEDICA Supercpecialty Hospital Kolkata | Sujay Krishna Maity et al | India/WB-IICB-036/2020 (EPI_ISL_661309) |
| National Virus Reference Laboratory | National Virus Reference Laboratory | Michael Carr et al | Ireland/CO-NVRL-75IRL32904/2020 (EPI_ISL_671872), Ireland/KE-NVRL-73IRL40299/2020 (EPI_ISL_578300) |
| Pathogen Genomics Center, National Institute of Infectious Diseases | Pathogen Genomics Center, National Institute of Infectious Diseases | Tsuyoshi Sekizuka et al | Japan/IC-0489/2020 (EPI_ISL_768698) |
| Institute for Medical Research, Infectious Disease Research Centre, National Institutes of Health, Ministry of Health Malaysia | Institute for Medical Research, Infectious Disease Research Centre, National Institutes of Health, Ministry of Health Malaysia | Suppiah J et al | Malaysia/IMR-WI194/2020 (EPI_ISL_718307) |
| Ministry of Health Hospitals | Institute of Health and Community Medicine | David Perera et al | Malaysia/UNIMAS-M4061/2020 (EPI_ISL_718165) |

| GISAID Acknowledgements | | | |
|---|---|---|---|
| Originating lab | Submitting lab | Authors | Samples |
| Dutch COVID-19 response team | National Institute for Public Health and the Environment (RIVM) | Adam Meijer et al | Netherlands/GE-RIVM-20154/2020 (EPI_ISL_723167) |
| LabPLUS | Institute of Environmental Science and Research (ESR) | Xiaoyun Ren et al | NewZealand/20CV0692/2020 (EPI_ISL_755627) |
| Ostfold Hospital Trust - Kalnes, Centre for Laboratory Medicine, Section for gene technology and infection serology | Norwegian Institute of Public Health, Department of Virology | Kathrine Stene-Johansen et al | Norway/3715/2020 (EPI_ISL_590987) |
| Haukeland University Hospital, Dept. of Microbiology | Norwegian Institute of Public Health, Department of Virology | Kathrine Stene-Johansen et al | Norway/5852/2020 (EPI_ISL_775336) |
| Laboratory for Respiratory Viruses, Cantacuzino National Military-Medical Institute for Research and Development | Cantacuzino Institute | M.Lazar et al | Romania/284213/2020 (EPI_ISL_455468) |
| HELIX LLC | WHO National Influenza Centre Russian Federation | Andrey Komissarov et al | Russia/MOS-RII-MH9270S/2020 (EPI_ISL_733427), Russia/PSK-RII-MH6191S/2020 (EPI_ISL_733361) |
| Institute of Virology, Biomedical Research Center of the Slovak Academy of Sciences, Bratislava | Faculty of Natural Sciences, Comenius University, Bratislava | Viktória Hodorová et al | Slovakia/UKBA-403/2020 (EPI_ISL_718251) |
| Institute of Virology, Biomedical Research Center of the Slovak Academy of Sciences, Bratislava | Faculty of Natural Sciences, Comenius University, Bratislava | Kristína Boršová et al | Slovakia/UKBA-404/2020 (EPI_ISL_718252), Slovakia/UKBA-604/2020 (EPI_ISL_788982) |
| NHLS-IALCH | KRISP, KZN Research Innovation and Sequencing Platform | Giandhari J et al | SouthAfrica/KRISP-DG01842216/2020 (EPI_ISL_736947) |
| NHLS Universitas Academic | UFS Virology | PA Bester et al | SouthAfrica/UFS-VIRO-NGS-35/2020 (EPI_ISL_682342) |

| GISAID Acknowledgements | | | |
|---|---|---|---|
| Originating lab | Submitting lab | Authors | Samples |
| Division of Emerging Infectious Diseases, Bureau of Infectious Diseases Diagnosis Control, Korea Disease Control and Prevention Agency | Division of Emerging Infectious Diseases, Bureau of Infectious Diseases Diagnosis Control, Korea Disease Control and Prevention Agency | Ae Kyung Park et al | SouthKorea/KDCA0108/2020 (EPI_ISL_747344) |
| Klinsisk mikrobiologi Linkoping | The Public Health Agency of Sweden | Department of Microbiology et al | Sweden/20-53002/2020 (EPI_ISL_661285) |
| Viollier AG | Department of Biosystems Science and Engineering, ETH Zürich | Christian Beisel et al | Switzerland/BL-ETHZ-230022/2020 (EPI_ISL_516565), Switzerland/BL-ETHZ-270140/2020 (EPI_ISL_541451) |
| University Hospital Basel, Clinical Virology | University Hospital Basel, Clinical Bacteriology | Madlen Stange et al | Switzerland/BS-42376604/2020 (EPI_ISL_581933) |
| Viollier AG | Department of Biosystems Science and Engineering, ETH Zürich | Chaoran Chen et al | Switzerland/SG-ETHZ-430703/2020 (EPI_ISL_796464), Switzerland/ZH-ETHZ-410010/2020 (EPI_ISL_737518) |
| 1-Laboratory of Microbiology, National Reference Lab, Charles Nicolle Hospital; 2-University of Tunis ElManar, Faculty of Medicine of Tunis, LR99ES09, Tunis, Tunisia | 1-Clinical and Experimental Pharmacology Lab, LR16SP02, National Center of Pharmacovigilance, University of Tunis El Manar, Tunis, Tunisia. 2-Neurodegenerative diseases and psychiatric troubles, LR18SP03, Razi Hospital, University of Tunis El Manar, Tunis, Tunisia. 3-Ministry of Health, National Observatory of New and Emerging Diseases, 1006, Tunis, Tunisia | Ilhem Boutiba-Ben Boubaker et al | Tunisia/3942/2020 (EPI_ISL_699657), Tunisia/61627/2020 (EPI_ISL_707698) |
| Uganda Central Public Health Lab and Uganda Virus Research Institute | MRC/UVRI & LSHTM Uganda Research Unit | Matthew Cotten et al | Uganda/UG089/2020 (EPI_ISL_738000) |

| GISAID Acknowledgements | | | |
|---|---|---|---|
| Originating lab | Submitting lab | Authors | Samples |
| Helix/Illumina | Genomics and Discovery, Respiratory Viruses Branch, Division of Viral Diseases, Centers for Disease Control and Prevention | Peter W. Cook et al | USA/AL-CDC-STM-034/2020 (EPI_ISL_802654) |
| Massachusetts General Hospital | Infectious Disease Program, Broad Institute of Harvard and MIT | Lemieux et al | USA/MA-MGH-03435/2020 (EPI_ISL_765837) |
| Utah Public Health Laboratory | Utah Public Health Laboratory | Erin Young et al | USA/NH-UPHL-2012558654/2020 (EPI_ISL_738357) |
| Wadsworth Center, New York State Department.of Health | Wadsworth Center, New York State Department.of Health | Kirsten St. George et al | USA/NY-Wadsworth-222417-01/2020 (EPI_ISL_765509) |
| Houston Methodist Hospital | Houston Methodist Hospital | S. Wesley Long et al | USA/TX-HMH-5181/2020 (EPI_ISL_544580) |
| Wales Specialist Virology Centre Sequencing lab: Pathogen Genomics Unit | COVID-19 Genomics UK (COG-UK) Consortium | Catherine Moore et al | Wales/PHWC-16C03E/2020 (EPI_ISL_572878) |
| National Institute for Communicable Disease Control and Prevention (ICDC) Chinese Center for Disease Control and Prevention (China CDC) | National Institute for Communicable Disease Control and Prevention (ICDC) Chinese Center for Disease Control and Prevention (China CDC) | Zhang et al | Wuhan/Hu-1/2019 (EPI_ISL_402125) |